\documentclass[aps,prx,twocolumn,superscriptaddress,nobibnotes]{revtex4-2}

\usepackage{amsmath,amssymb}
\usepackage{graphicx}
\usepackage{color} 
\usepackage[T1]{fontenc} 
\usepackage{bm}

\usepackage{xspace}  
\usepackage{multirow}
\usepackage{dcolumn} 
\usepackage{booktabs}

\usepackage{xspace}  
 
\newcommand{\etal}{\textit{et al.}\xspace}

\begin{document}

\title{How long can a non-spherical quantum object remain standing? \\- a fundamental quantum question -}

\newcommand{\ariken}{      \affiliation{RIKEN Nishina Center, 2-1 Hirosawa, Wako, Saitama 351-0198, Japan}}
\newcommand{\aut}{         \affiliation{Department of Physics, The University of Tokyo, 7-3-1 Hongo, Bunkyo, Tokyo 113-0033, Japan}}
\newcommand{\atsuku}{        \affiliation{Center for Computational Sciences, University of Tsukuba, 1-1-1 Tennodai, Tsukuba, Ibaraki 305-8577, Japan}}

\newcommand{\aemp}{\email{otsuka@phys.s.u-tokyo.ac.jp}}  
 
\author{Takaharu Otsuka$^*$}     \aut \ariken \aemp
\author{Yusuke Tsunoda}       \atsuku

\date{\today}

\begin{abstract}   
\bf{\noindent
{\it - Background} An isolated quantum system generally exhibits rotational symmetry, i.e., conserved spin (angular momentum) in its eigenstates.  Non-spherical quantum objects such as many of molecules and atomic nuclei are not exceptions.  However, these objects are not rotationally invariant by definition.  Such an object therefore restores, as a consequence of the action of Hamiltonian, the rotational symmetry by superposing states of the same object orienting in different directions, where each component represents one direction. For spin-0, the superposition is made isotropically, for instance.  Stationary eigenstates of a non-spherical object are thus formed, being time-independent.\\  
  {\it - Finding} We show the time evolution of an individual component of this superposition: 
this component remains almost unchanged for finite time, called standing time.  This implies that if the object is found to be in this component, it basically remains so for the standing time. This feature is shown to be relevant in a variety of objects, such as atomic nuclei, polymers (proteins), and electron drops in atoms.\\
  {\it - Nuclear shapes}  
Atomic nuclei are isolated quantum objects.   
The shapes of many nuclei are ellipsoids with variations, and their stationary eigenstates are indeed superpositions of the same ellipsoid orientated in different directions. The ``viewing'' of the ellipsoidal shape is not straightforward, because this ellipsoid is not at rest as in the classical case. Although physical observables related to shapes, e.g., quadrupole moments, have been measured, a ``snapshot'' of a nucleus is highly desired as a direct and complementary information. 
Recent experimental approaches with Relativistic Heavy-ion Collision (RHC) are promising for taking such a snapshot.  While its feasibility may be under investigations, the present work depicts that the standing time, $\sim$some 10$^{-23}$ sec for typical ellipsoidal nuclei, is much longer than the time scale of  RHC, $\sim$some 10$^{-25}$ sec.  This implies that an ellipsoidal nucleus remains practically unchanged for this critical period, despite that it is a part of stationary eigenstate with rotational symmetry.  
The snapshot experiments with RHC is thus shown to be feasible.\\   
{\it - Generality} As the standing time will be intimately related, through a relation like the energy-time  uncertainty relation, to the energy scales involved, we can also explore this concept and its applications in a variety  of physical systems.\\ 
{\it - Nuclear reactions} The standing time is close to the tunneling time through the barrier in sub-barrier fusion reactions, consistent with an existing theoretical model. The hot fusion reactions for the synthesis of superheavy elements are facilitated by anticipated longer standing time. The present work is consistent with the time scale of fission reactions and is supportive for simpler modelings on them. A variable standing time seems to enhance $\alpha$ decay/emission.\\ 
{\it - Protein and electron}  Beside nuclear physics cases, similar studies are possible for other systems with different values of the standing time due to different energy scales. Proteins share several basic properties with nuclei, with longer standing times predicted, for instance, $\sim$ 2 nano sec. An intriguing property of an electron cooled down and placed in an atomic Coulomb potential is discussed, referring to a narrower energy range with the Coulomb potential.\\
{\it - Summarizing remark} This novel feature is unveiled by identifying symmetry-breaking (e.g., ellipsoid shape) states arising from stationary states with symmetry, for the period of standing time, within the same framework of Schr\"odinger equation. More cases in wider disciplines are of interest.
}
 \end{abstract}   

\maketitle

\section{Introduction}


\begin{figure*}[tb]
  \centering
\includegraphics[width=18cm]{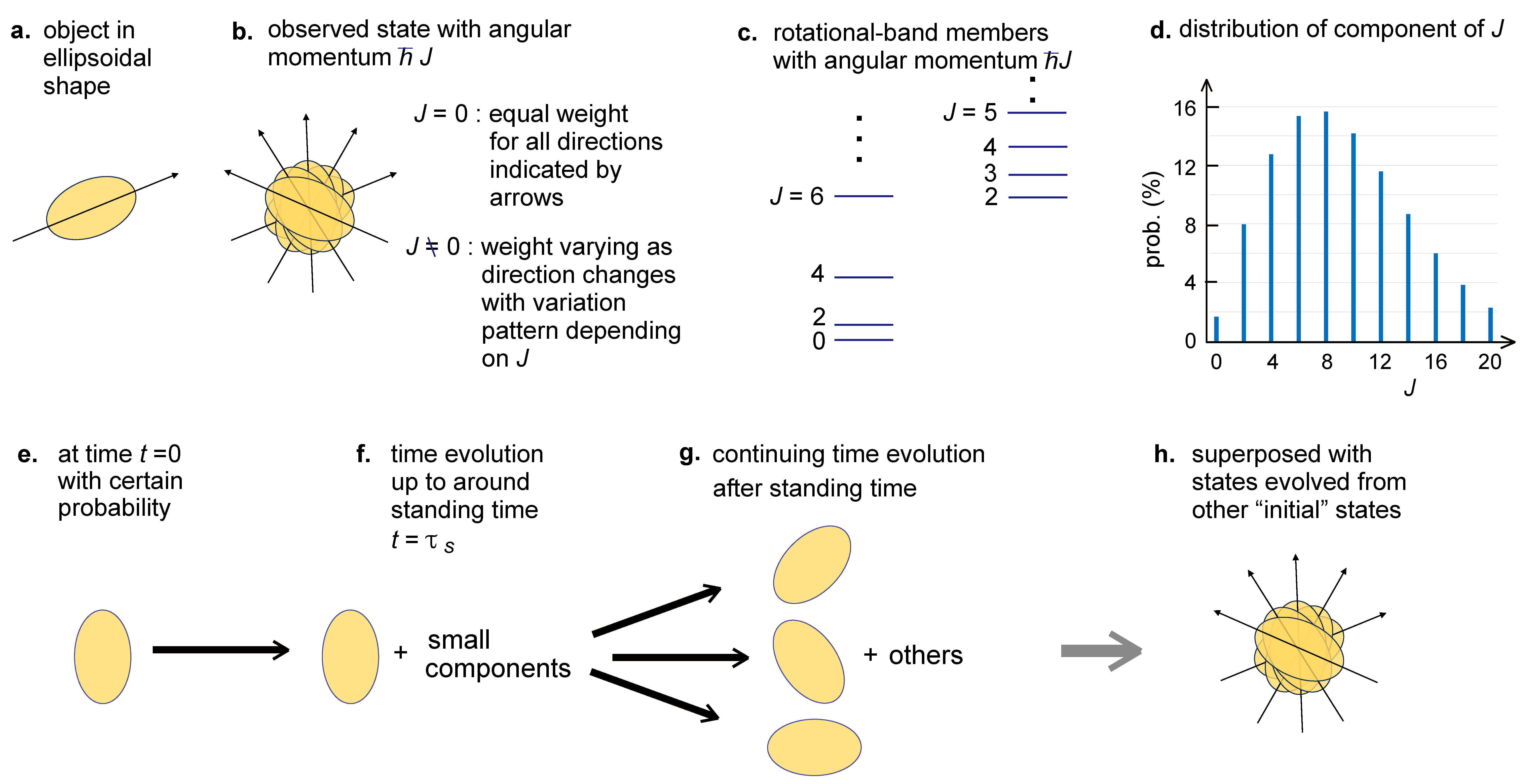}
    \caption{ {\bf Schematic illustration of non-spherical (ellipsoidal) state, and its time evolution and rotational mode. }  
  {\bf a}, Non-spherical object modeled by an ellipsoidal. The arrow denotes the longest principal axis. This object is described by intrinsic state, and can be an atomic nucleus.  
  {\bf b}, Quantum states with angular momentum $\hbar J$ created by superposing the ellipsoidal states in different directions.
  {\bf c}, Excitation energies of members of rotational bands. 
  {\bf d}, Typical probabilities of members of angular momentum, $J$, in the rotational band built on the ground state (c.f. left band in panel {\bf c}) of the nucleus $^{166}$Er.  
  The height of the bin represents the probability $|f_J|^2$ in eq.~(\ref{eq:phi}) normalized within the band.  
  {\bf e-g}, Schematic illustration of the time evolution of the nucleus in an ellipsoidal shape.
  {\bf e}, The ellipsoidal nucleus is pointing in a certain direction. {\bf f}, Time-evolved states 
  until standing time.  {\bf g}, Time-evolved states well after standing time.  {\bf h}, Stationary state as the  superposition of states initiated in all directions, being identical to panel {\bf b}.} 
  \label{fig:rot_state}  
\end{figure*}

The shape of any object is of natural and fundamental interest as well as of practical importance, regardless of the object size. The shape is a concept attached to the object: the shape can be viewed well when it is at rest, relative to the observer. 
This is not a major problem in classical mechanics, but can be 
with quantum mechanics.  In quantum mechanics, an isolated system must fulfill  
rotational symmetry, which is brought in by the Hamiltonian that is supposed to be rotationally invariant because of the perfect isolation (no external field).

A non-spherical quantum object is considered in this article.  Although the shape is general here, for the sake of simplicity, we consider ellipsoidal objects as shown in Fig.~\ref{fig:rot_state}{\bf a}.  
The arrow in the figure indicates the direction of the longest principal axis.  The wave function describing the object in Fig.~\ref{fig:rot_state}{\bf a} is called the 
intrinsic state.
In the eigenstate of the Hamiltonian, the Hamiltonian mixes (superposes) the same object 
state (intrinsic state) oriented in different directions, as schematically shown in Fig.~\ref{fig:rot_state}{\bf b}.  
We shall discuss the relation between the state in Fig.~\ref{fig:rot_state}{\bf a} and the state in Fig.~\ref{fig:rot_state}{\bf b} focusing on their time evolution properties,  
in a quantum mechanical description. 
A novel feature and wide relations to various phenomena will be presented.

The precise clarification of the shape like Fig.~\ref{fig:rot_state}{\bf a} is a major subject of physics in general, independently of object type or size. But the quantum reality is like Fig.~\ref{fig:rot_state}{\bf b}.  Although physical observables of the eigenstate shown in Fig.~\ref{fig:rot_state}{\bf b} carry useful information resulting from the intrinsic state in Fig.~\ref{fig:rot_state}{\bf a}, the obtained information is indirect and can be model/theory dependent. 
It is important to find a way to extract the shape information directly from the state in Fig.~\ref{fig:rot_state}{\bf a}.
This paper presents whether and how this can be possible, as well as applications to other phenomena.

The subject and discussion are general, but we use ellipsoidal deformed atomic nuclei in order to illustrate the time-evolution process in a concrete fashion.   
Atomic nuclei are on the scale of quantum mechanics, being the possibly tiniest objects with shapes, 
and exhibit rotational bands (Sec.~\ref{band}).  
The time evolution (Sec.~\ref{evolution}) and the standing time (Sec.~\ref{standing}) are described in simple and general terms.

The shape of the atomic nucleus has been intensely studied over many decades since the 1950s\cite{rainwater1950,bohr_1952,bohr_nobel,bohr_mottelson_book2,casten_book,heyde_book,ring_schuck_book,deshalit_feshbach_book}.  These studies 
have focused on shape-sensitive observables with huge success for instance by electric quadrupole moments \cite{stone_2005} and transitions\cite{pritychenko_2016}, multiple Coulomb excitations\cite{cline_1986}, {\it etc.}, but nuclear shapes remain to be elucidated more precisely, including spin-0 ground state 
of even-even nuclei.  
It is desired to see the shapes like Fig.~\ref{fig:rot_state}{\bf a}, but this is fundamentally difficult because, as already stated from general quantum mechanical viewpoint,  the nucleus does not remain at rest and is taking different orientations.  Although this differs from the classical image of rotation, it is a well-accepted feature (e.g., \cite{ring_schuck_book} and more detailed illustrations focused on this point in \cite{otsuka_2025,otsuka_2025a,otsuka_2026}). So, if the shape can be observed, it has to be a 
snapshot\cite{STAR_2024}.  While its general feasibility may be under further investigations, this work indicates its feasibility in terms of the standing time (Sec.~\ref{concrete}). 
The standing time is further shown to be related 
to sub-barrier fusion\cite{hagino_2012,hagino_2022} (Sec.~\ref{fusion}), fission in general\cite{vandenbosch_book,krappe_book,schunk_2022} (Sec.~\ref{fission}) and hot fusion for superheavy element synthesis\cite{karpov_2013,erler_2012,smits_2024,oganessian_2024} (Sec.~\ref{super}), which are typical  short-time behaviors of quantum systems.  
We make remarks on $\alpha$ decay/emission (Sec.~\ref{fission}).

Some discussions on the restoration of broken symmetry are presented (Sec.~\ref{broken}), pointing out the relevance of energy scale. The studies are then extended beyond nuclear physics.  One interesting case is found with polymers: we estimate the standing time for a heavy molecule, protein, with a much longer standing time due to much smaller energy scale (Sec.~\ref{protein}).
Another intriguing case is an electron drop placed in a hydrogen-like potential.  The standing time can be considered for the location of this electron (Sec.~\ref{electron}).

A summary and remarks are presented at the end. 

Regarding the snapshot, a new experimental approach has emerged, where ultra-high-speed snapshots\cite{STAR_2024,giacalone_2022,bally_2022,jia_2022,dimri_2023, giacalone_2023,Ryssens_Giacalone_2023,adamczyk_2015,alice_2018,sirunyan_2019,aad_2020,abdallah_2022,alice_2022,aad_2023,duguet_2025} can be taken 
with a very short timescale like 10$^{-25}$ sec, by using the relativistic heavy-ion collision (RHC).  
To demonstrate the RHC approach, an argument for the time scale was made\cite{STAR_2024}, 
by comparing to a time scale estimated for fluctuating angle of the axis of a deformed nucleus \cite{nakatsukasa_2016}.
Such arguments, however, sparked a commentary warning paper\cite{doba_2025}.

\section{Rotational Band of Nuclei in Ellipsoidal Shapes. \label{band}}
In a conventional semi-classical picture, a rotational band is created by an ellipsoid rotating with spin or angular momentum $\hbar J$.  The ellipsoid here is nothing but the nucleus with an ellipsoidal shape (see Fig.~\ref{fig:rot_state}{\bf a}).
By assuming this motion as a free rotation of an axially-symmetric object, its quantum mechanical treatment yields excitation energies $\propto J(J+1)$ with $J$=0, 2, 4, 6, ... for the energetically lowest (i.e., ground) band of the nucleus (see the left part of Fig.~\ref{fig:rot_state}{\bf c}).  The parity is assumed to be positive, but is omitted for simplicity hereafter.   Even values of integers are taken for $J$, partly because we consider nuclei with even numbers of protons and neutrons.  Odd values appear in the other band in Fig.~\ref{fig:rot_state}{\bf c}. Details of such well-known features are not relevant to this article, and are not mentioned.

Beautiful rotational bands appear also in a fully quantum mechanical description of many-body systems, without classical spinning \cite{otsuka_2025}. The actual picture of rotational excitation energy differs from that in conventional semi-classical pictures, as presented in detail up to recent findings in \cite{otsuka_2025} (or its digest edition\cite{otsuka_2025a}) and \cite{otsuka_2026}.  
In such fully quantum descriptions, the intrinsic state, $\phi_i$, is introduced as a quantum many-body state representing a nucleus in an ellipsoidal shape like the one in Fig.~\ref{fig:rot_state}{\bf a}.  A simple image of $\phi_i$ may be a Hartree-Fock state with a deformed mean potential, but actual $\phi_i$ is of more complex and richer structure, carrying various correlation effects from nuclear forces.     
A quantum mechanical definition of rotational band can be given by that the band member with angular momentum $J$ is created by superposing the states obtained by rotating $\phi_i$ to various directions as schematically shown in Fig.~\ref{fig:rot_state}{\bf b}.  This is a natural definition in the sense that these band members are created from the same seed, i.e., the common intrinsic state. 
The wave functions of the rotated states are linearly added in the superposition, where the weighting factors have different dependences on the Euler angles for different $J$ and other additional quantum numbers. Mathematically, the whole process is called angular-momentum projection, which has been well established by means of Wigner's $D$-function\cite{edmonds} (for details, e.g., see {\it Appendix} or \cite{ring_schuck_book}), and it is well accepted\cite{ring_schuck_book,otsuka_2025} that this scheme can provide with eigenstates of the rotational bands of current interest in a good approximation.  

For instance, the state with $J$=0 is represented by wave function $\psi_0 ({\bm x})$, where ${\bm x}$ collectively denotes all possible coordinates of all active nucleons.  It can be constructed by the superposition of $\phi_i$ rotated to all angles with equal weight, as schematically shown in Fig.~\ref{fig:rot_state}{\bf b}.  The angles are three Euler angles denoted by ${\bm \omega}$.   As these angles are continuous variables, the superposition is expressed by means of an integral with respect to Euler angles,
\begin{equation}
\psi_0 ({\bm x}) \,=\,\frac{1}{8 \pi^2 {\cal N}} \, \int  d{\bm \omega } \big\{ D^0_{0,0} ({\bm \omega}) \big\}^*\,  \hat{R}({\bm \omega})\phi_i ({\bm x}),
\label{eq:rot_phi}
\end{equation}
where ${\cal N}$ denotes a normalization constant, $\hat{R}({\bm \omega})$ is the rotation operator by Euler angles ${\bm \omega}$.
The function $D^0_{0,0}$ is a special case of the $D$ function for $J$=0, and its value is unity.
In this way, the intrinsic state $\phi_i ({\bm x})$ and the eigenstate $\psi_0 ({\bm x})$ are related, and other band members are generated in a similar way adjusted for individual $J$ value.

The feature presented above can be inversely considered.  Because the angular-momentum projection is a kind of filtering, the wave function of $\phi_i$ should contain eigen wave function, $\psi_J ({\bm x})$, of the band member with angular momentum $J$; otherwise $\psi_J ({\bm x})$ cannot be filtered out.  We can then write down as,
\begin{equation}
\phi_i ({\bm x}) \,=\, f_0 \, \psi_0 ({\bm x}) \,+\, f_2 \, \psi_2 ({\bm x}) +  f_4 \, \psi_4 ({\bm x}) + \cdots 
\label{eq:phi}
\end{equation}   
where $f_J$ is amplitude normalized so that $\sum_J |f_J|^2 = 1$.  Here, $\psi_J ({\bm x})$ is normalized, and includes all subcomponents of magnetic quantum number $J_z$ ranging from $-J$ to $+J$ with proper weights.  The states in eq.~(\ref{eq:phi}) are highly complex multi-nucleon states, but such practical details are irrelevant in the following discussions.   Figure \ref{fig:rot_state}{\bf d} displays a typical pattern of $|f_J|^2$.  
The value of $|f_J|^2$ increases as $J$ increases from 0, and decreases after passing its peak. This  general property is explained here. In the intrinsic state, couplings of $\psi_J$'s of different $J$'s naturally occur  due to (intrinsic) deformed fields, and provide more binding energies.  This mechanism increases $|f_J|^2$ up to a certain $J$.  However, beyond a certain $J$ value, the energy of $\psi_J$, basically proportional to $J(J+1)$,  becomes so high, and cancels the energy gain due to the mixing.  This change of the balance leads to the decrease of $f_J$ for higher $J$'s. 

The definition of the state denoted by $J$ (see eq.~(\ref{eq:phi})) remains valid for general cases involving so-called $K$ quantum number (see {\it Appendix} for details), where $J$ implicitly includes $K$, e.g., ($J=2, K=0$) or ($J=2, K=2$) as different states. We note that this practical conservation of $K$ quantum number in actual deformed nuclei has been verified by means of realistic Configuration Interaction (CI) calculations \cite{otsuka_2025}.    
\\

\section{Time Evolution of Intrinsic State \label{evolution}}   

The nucleus is supposed to be in its stationary ground state (see Fig.~\ref{fig:rot_state}{\bf b}), and that at a certain time denoted by $t$=0, this ground state is in its component $\phi_i$ shown in Fig.~\ref{fig:rot_state}{\bf e}. This happens with a certain probability, as a realization of the probability principle in quantum mechanics proposed by Born\cite{max_born_lecture}, but may not have attracted attention in the study of rotational modes.  
The state $\phi_i$ is not an eigenstate, and it must change to other states as time goes by (see Fig.~\ref{fig:rot_state}{\bf f},{\bf g}).  

The time evolution of quantum states is driven by the Hamiltonian, $H$, following the time dependent Schr\"odinger equation,
\begin{equation}
i \, \hbar \, \frac{\partial}{\partial t} \phi ({\bm x}; t) \,=\, H \phi ({\bm x}; t)  ,
\label{eq:Sch_time}
\end{equation}
where $\phi({\bm x}; t)$ is the wave function at time $t$. 
It is then assumed that among various components of the ground state, $\phi_i$ comes in at $t=0$, 
\begin{equation}
\phi({\bm x}; t=0)=\phi_i ({\bm x}),
\label{eq:Sch_time_2}
\end{equation}
which gives an initial condition to eq.~(\ref{eq:Sch_time}). 
Some additional process of interest, such as taking snapshot or initiating fusion, is supposed to start at $t=0$ with $\phi_i$.  
The solution of eq.~(\ref{eq:Sch_time}) then gives  the time evolution of the ellipsoidal-nucleus part, in parallel with the additional process.  The solution of eq.~(\ref{eq:Sch_time}) is obtained by applying the Hamiltonian as the time evolution operator, and the outcome is straightforward, because $\phi_i ({\bm x})$ is given in eq.~(\ref{eq:phi}) where 
$\psi_J ({\bm x})$ is an eigenstate of $H$ with its eigenvalue, $E_J$. The time-dependent wave function is therefore expressed as,
\begin{equation}
\phi ({\bm x}; t) \,=\, \sum_J \, f_J \, \psi_J ({\bm x}) \, e^{-i (E_J - \bar{E}) t / \hbar}.
\label{eq:wav_time}
\end{equation}
where $\bar{E}$ stands for the origin of the energy coordinate. The value of $\bar{E}$ does not change the essential consequence, and can be conveniently chosen. 
The choice of $\phi_i$ at $t=0$ is made without losing generality, as $\phi_i$ can be replaced with an intrinsic state in any direction.

The intrinsic state $\phi_i ({\bm x})$ evolves into various states as time $t$ goes by (see Fig.~\ref{fig:rot_state}{\bf f},{\bf g}).  The time-dependent state $\phi ({\bm x}; t)$ then remains to be $\phi_i ({\bm x})$ with the probability amplitude given by an overlap, 
\begin{eqnarray}
r(t) &=& \int d{\bm x}  \langle \phi ({\bm x}; t=0) \,| \phi ({\bm x}; t) \rangle \nonumber \\
 &=& \sum_J \, |f_J|^2  \,  e^{-i (E_J-\bar{E}) t / \hbar}  \nonumber \\
 &=& \sum_J \, |f_J|^2  \, \{ {\rm cos}((E_J-\bar{E}) t / \hbar) - i \,{\rm sin}((E_J-\bar{E}) t / \hbar) \},  \nonumber \\
\label{eq:ov_time}
\end{eqnarray}
where the integral is generally multi-dimensional as ${\bm x}$ implies all coordinates for all active particles.  Such integral calculations are not necessary if we use energy eigenvalues.
While $r(t=0)$=1, $|r(t=0)|$ generally becomes smaller than unity for $t$> 0, due to the loss of coherence in the summation over the eigenstates.  It is commented that state $\psi_J$ contains components with different values of $J_z$, but they have the same energy eigenvalues, following the same time evolution. For this reason, the $J_z$ quantum number does not appear in eq.~(\ref{eq:ov_time}).

\begin{figure*}[bt]
  \centering
\includegraphics[width=17cm]{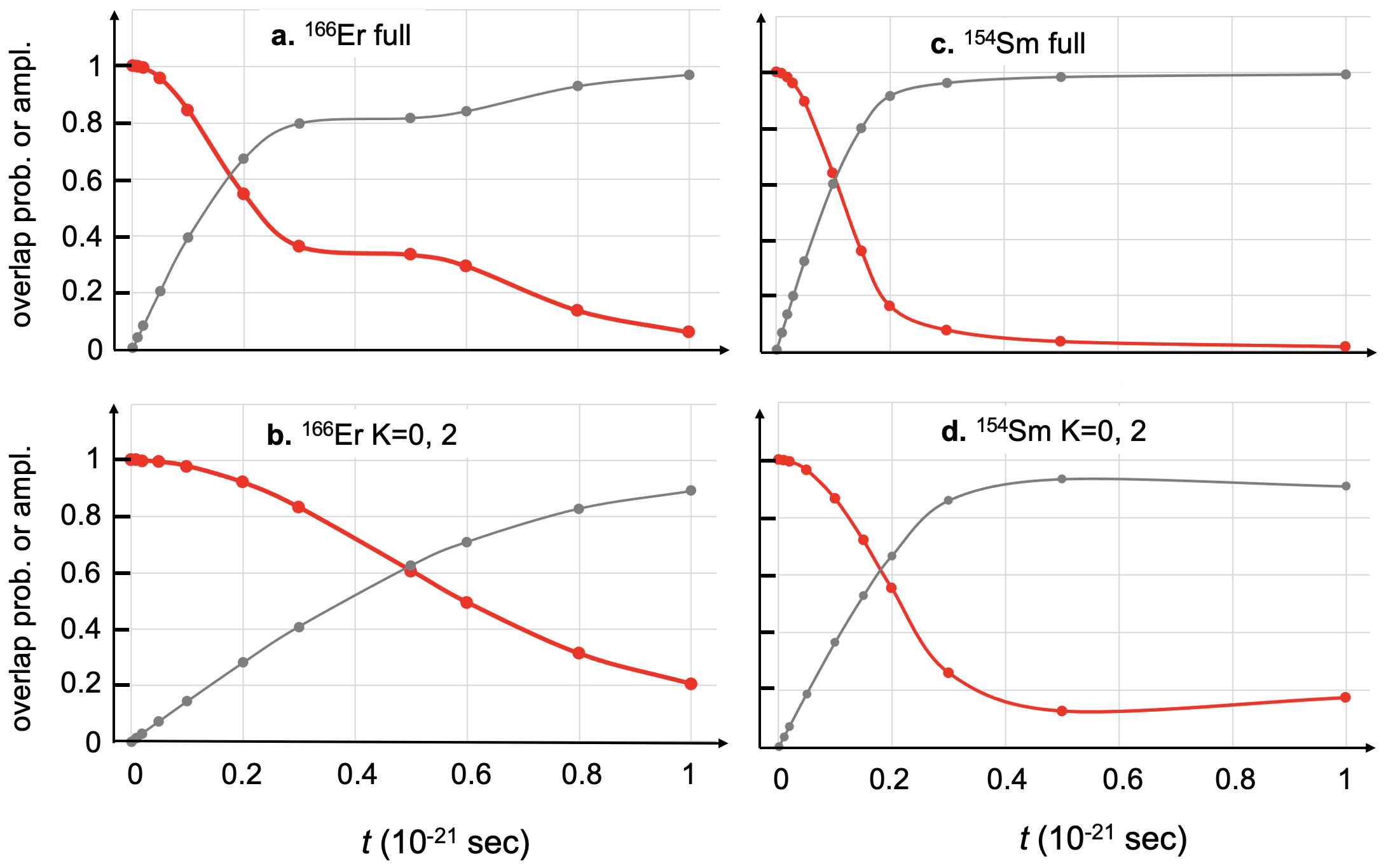}   
    \caption{ {\bf Time evolution of standing probability of ellipsoid. }  
    Red lines denote the squared magnitude of the overlap function, $|r(t)|^2$ (see eq.~(\ref{eq:ov_time})) as a function of time in the unit of 10$^{-21}$ sec.
    Gray lines similarly indicate the magnitude of the amplitude of the remaining orthogonal state.  
  {\bf a, b} are for $^{166}$Er, while {\bf c, d} for $^{154}$Sm.       
  In {\bf a, c}, all bands up to $K$=20 are included, while in {\bf b, d} $K$=0 and 2 bands. } 
  \label{fig:overlap_ampl}  
\end{figure*}  

\section{Standing Time of Ellipsoidal State \label{standing}} 

The overlap amplitude $r(t)$ remains near unity for time $t$ close to 0 (see Fig.~\ref{fig:rot_state}{\bf f}).  If $r(t)$ is close enough to unity, the physical quantities obtained for $\phi (x; t)$ are dominated by the value for $\phi (x; t=0) = \phi_i$, and the measurements of these quantities are interpreted as those for the intrinsic state, with this understanding.    

It is of extreme interest in what pace the overlap amplitude $r(t)$ deviates from unity.  We set a minimum value, $r_{min}$ for $r(t)$: if $r_{min}$= 0.98, the remaining component of $\phi (x; t)$ orthogonal to $\phi_i$ has amplitude with its magnitude about 0.20.  Once this magnitude becomes greater than this value, the orthogonal component may start to produce notable contributions in certain circumstances.  Of course, this criterion depends on the cases, and the present one may be too much on the safe side. Nevertheless, we tentatively 
adopt it for the sake of definiteness.
The time duration corresponding to this boundary is hereafter called the {\bf standing time}, denoted by $\tau_s$.  So, we suppose that within $t < \tau_s$, the properties of $\phi (x; t)$ are basically identical to those of $\phi_i$, implying that we can look into the features of $\phi_i$.  We note that the ``standing'' of the standing time does not mean being upright but implies being unchanged
(to a good extent). 
 
We estimate how $r(t)$ changes for $t$ not so large, supposing a probable distribution of $|f_J|^2$, e.g.,  Fig.~\ref{fig:rot_state}{\bf d}.   We set $\bar{E} \sim E_J$ of the maximum value $|f_J|^2$, and introduce $\Delta E_J  \,\equiv\,  E_J-\bar{E}$. 
The contributions from the states of $\Delta E_J \sim 0$ are characterized by the cosine part $\sim$ 1, whereas the sine part $\sim$ 0 in eq.~(\ref{eq:ov_time}).   
For the states of $\Delta E_J \sim 0$, the sum of their $|f_J|^2$'s is denoted by $\alpha$.  
Their contribution to $r(t)$ is then approximated by $\alpha$.

The representative value of $\Delta E_J$ for the rest of the states is denoted by $\Delta E_r$. The main contribution to $r(t)$ from the rest of states is approximated by the cosine part with $\Delta E_r$ in eq.~(\ref{eq:ov_time}), with the sine part neglected. The relevant $|f_J|^2$ factor is (1 - $\alpha$).
We thus come up with
\begin{equation}
r(t=\tau_s) \, \sim \, \alpha \,+ \, (1-\alpha)  \, {\rm cos}(\Delta E_r  \, \tau_s / \hbar) = r_{min}. 
\label{eq:St_time_est1}
\end{equation}
This equation leads, for $\Delta E_r$ in MeV, to
\begin{equation}
\tau_s \, \sim \, 6.6 \, \eta \, \frac{1}{\Delta E_r} \, 10^{-22} \, {\rm sec}, 
\label{eq:St_time_est2}
\end{equation}
where $\hbar$=6.6 $\times$ 10$^{-22}$ MeV$\cdot$sec is substituted, and $\eta$ is introduced as a parameter defined by ${\rm cos} \,\eta\,= (r_{min}-\alpha)/(1-\alpha)$, for brevity.
By taking $r_{min}$=0.98 as discussed already, and by assuming $\alpha$=1/2 (1/4) as a representative possible value, $\eta$=0.29 (0.24) is obtained. 
The explicit expression for ${\rm cos} \,\eta$ indicates weak dependences on $r_{min}$ and $\alpha$. For example,  $\eta$ increases by about 60\% for a rather extreme value, $r_{min}$=0.95 (10 \% loss of the probability).

In more realistic calculations, more than one band are generated from $\phi_i$, and are distinguished by so-called $K$ quantum number, e.g., the band on the right-hand side in Fig.~\ref{fig:rot_state}{\bf c}.  Those bands are included in numerical calculations below.  Eigenstates are labelled by $K$ besides $J$ and $J_z$.  
The representative energy spread ${\Delta E_r}$ then appears to be 2-4 MeV, with which eq.~(\ref{eq:St_time_est2}) gives an estimate that the standing time, $\tau_s$, is somewhat shorter than 10$^{-22}$ sec.
Although this is a very simple na\"ive estimate, it seems to illuminate a reasonable order of magnitude.

Equation (\ref{eq:St_time_est2}) can be rewritten as $\tau_s \, \Delta E_r \sim \eta \, \hbar$.  Remembering $0 < \eta \ll 1$, this looks like the time-energy uncertainty principle.  This kind of relations generally arise when the time dependence of wave functions is looked into (recall eq.~(\ref{eq:St_time_est2})).  The energy spread ${\Delta E_r}$ intuitively stands for the difference between the representative energy of the states most contributing to disappearance of $\phi_i$ and the representative energy of the states most contained in $\phi_i$.  So, it implies the energy of primary changes at the initial stage.

We note, for the sake of confirmation, that even if the nucleus is completely isolated with no additional process, the time evolution discussed in Secs.~\ref{evolution} and \ref{standing} progresses but the stationary state is kept the same, owing to the proper superposition of intrinsic states in all orientations (see Fig.~\ref{fig:rot_state}{\bf h}). Once an additional process (e.g, collision, fusion, fission, {\it etc}) starts with an intrinsic state $\phi_i$ arising with   quantum-mechanical probability (Fig.~\ref{fig:rot_state}{\bf e}), the time evolution of this $\phi_i$ matters. The time-evolved state remains very close to $\phi_i$ for a certain time (Fig.~\ref{fig:rot_state}{\bf f}), before it changes to other orthogonal states much afterwards (Fig.~\ref{fig:rot_state}{\bf g}).  This time duration is called the standing time, and, for instance, the intrinsic shape may be ``directly seen'' during the standing time.
\\

\section{Concrete Time Evolution for Heavy Nuclei \label{concrete}} 

We now present some concrete calculation using intrinsic states given by state-of-the-art Configuration Interaction (CI) calculations\cite{otsuka_2025} for heavy deformed nuclei, suitable for the present study. 
Specifically, as the intrinsic state $\phi_i$, we use the state $\xi_3$ thus obtained\cite{otsuka_2025} for describing properties of erbium-166 ($^{166}$Er, $Z$=68, $N$=98) nucleus.
This nucleus is chosen as it has been taken as a representative case of a nucleus with ellipsoidal deformed shape (see, for instance, \cite{bohr_nobel,bohr_mottelson_book2,otsuka_2025,otsuka_2025a,otsuka_2026}).    

The expansion in eq.~(\ref{eq:phi}) is carried out for $\xi_3$ including side bands, one of which is schematically shown in the right part of Fig.~\ref{fig:rot_state}{\bf c}.   
To be more technical, 
the band members of $J$=0-20 (only even integers) in the lowest $K$=0 band, those of $J$=2-20 in the $K$=2 band, and those of $J$=4-20 in the $K$=4 band are included.  In addition, the bands of $K$=6-20 (even integers) are included; the energies of band members are incorporated not individually but in average through the energies of $K$-projected intrinsic states for simplicity.  

Figure~\ref{fig:overlap_ampl}{\bf a} shows (red line) the calculated squared overlap amplitude, i.e., overlap probability,  $|r(t)|^2$, as a function of time $t$.  In the same panel, the magnitude of the amplitude of the remaining orthogonal state is shown by gray line.  The sum of red-line value and the square of gray-line value is always unity.

The probability of $\phi_i$ remains rather close to unity (>0.9) until 0.7$\times$10$^{-22}$ sec, and decreases gradually.  The standing time appears to be 0.3$\times$10$^{-22}$ sec for the lower bound of the probability, 0.96, fixed above ($r_{min}$=0.98).  
This value of standing time is consistent with the simple estimate given by eq.~(\ref{eq:St_time_est2}).  
Because of the gradual decrease, the standing time becomes longer 
if the criterion is made looser with a smaller value of $r_{min}$. The criterion depends on physical quantity of interest and its precision, and 
the presently taken value $r_{min}$=0.98 certainly lies on a strict side. The probability decreases up to  $t\sim$0.2$\times$10$^{-21}$ sec in a manner similar to cosine function.  The probability stays low afterwards because of lost coherence in eq.~(\ref{eq:wav_time}).  The probability may come back in principle, but will never come back to unity unless a very exceptional situation arises.  It is worth mentioning, though maybe trivial, that the probability does not decrease like an exponential, as this is not a decay.


\begin{figure}[tb]
  \centering
\includegraphics[width=8.5cm]{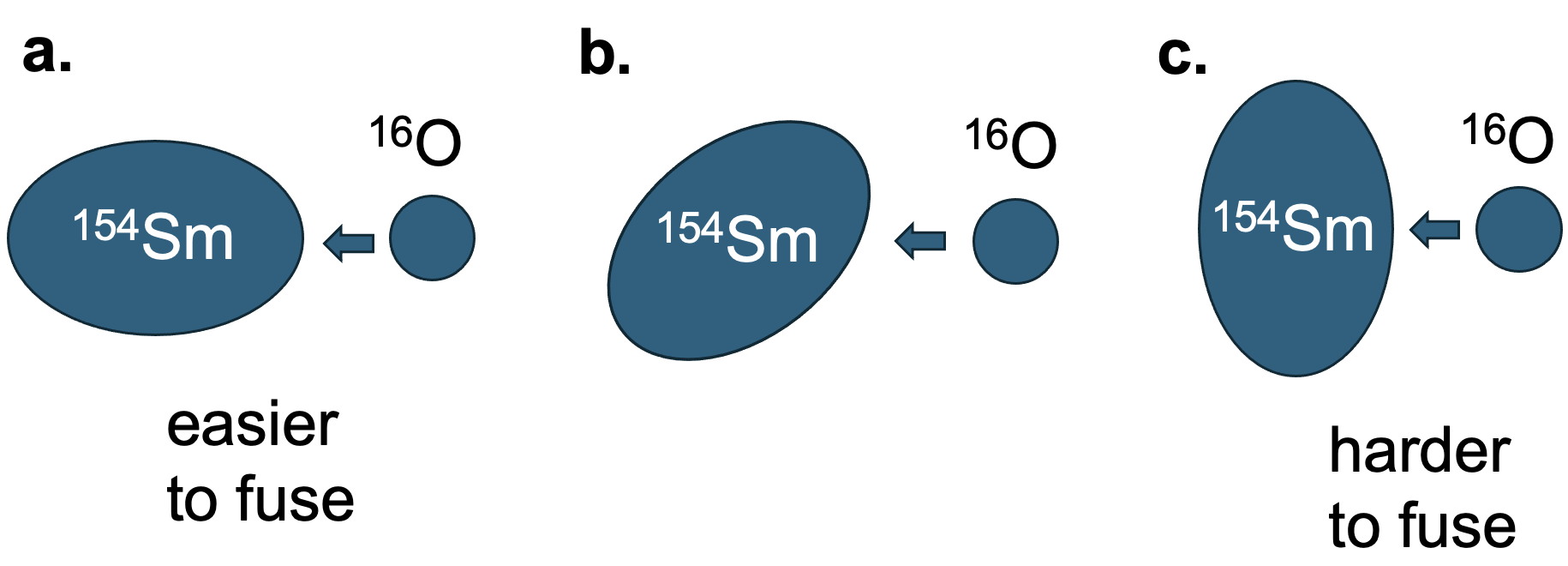}
    \caption{ {\bf Fusion raection. }  
    {\bf a.} $^{16}$O towards elongated edge,  
  {\bf b.} towards a tilted side, 
   {\bf c.} towards flat side.} 
  \label{fig:fusion}  
\end{figure}  

For a comparison, the calculation including only $K$=0 and 2 bands is shown in Fig.~\ref{fig:overlap_ampl}{\bf b}.   The time evolution proceeds more slowly, by a factor of about 2.5, than in Fig.~\ref{fig:overlap_ampl}{\bf a}. 
This is natural, because fewer states now participate in the time evolution.  It is of importance that the difference is as small as a factor of 2.5.   Thus, it does not matter too much which states of higher energies are included in the calculation as long as the basic feature is concerned.

Figure~\ref{fig:overlap_ampl}{\bf c} shows (red line) the same calculation for samarium-154 ($^{154}$Sm, $Z$=62, $N$=92) nucleus\cite{otsuka_2025}. 
This nucleus has also been considered often as a good example in discussions on the ellipsoidal shape and the rotational modes, but is known to show different structure details as compared to $^{166}$Er, discussed above (see, for instance, \cite{otsuka_2025}). 

Time evolution is slightly faster than for $^{166}$Er.  Otherwise, the basic patterns are the same as the $^{166}$Er.  Figure~\ref{fig:overlap_ampl}{\bf d} corresponds to Fig.~\ref{fig:overlap_ampl}{\bf b}. 
 
The general arguments and numerical calculations both suggest that the standing time is some 10$^{-23}$ sec for the nuclei with mass number $A$(=$Z$+$N$)$\sim$160.  
To be more concrete, the standing time is shown to be 5 (3) $\times$10$^{-23}$ sec for $^{166}$Er ($^{154}$Sm) for the present strict criterion, $r_{min}$=0.98.
The criterion can be loosened as mentioned above.

We emphasize that the present work is about the time evolution but not about any angular motion of the nucleus in an ellipsoidal shape. We (can) avoid such angular motion, a classical image.
    
The time period during which the ellipsoid orientation remains almost in a fixed direction was mentioned in \cite{STAR_2024} to be 3$\times$(10$^{-21}$-10$^{-20}$) sec quoting the number in \cite{nakatsukasa_2016}, 
for the purpose of demonstrating the RHC experiment for the elucidation of nuclear shapes. This number is 100-1000 times larger than the present result.  It is expected from Fig.~\ref{fig:overlap_ampl} that before reaching $t$=3$\times$(10$^{-21}$-10$^{-20}$) sec, the overlap probability $r(t)$ becomes quite small in magnitude, and behaves irregularly due to lost coherence.  The result of \cite{nakatsukasa_2016} was obtained by applying the uncertainty principle of the angle and the angular momentum. Its relevance to the present case is not clear. Since the nucleus in ellipsoidal shape is a compact object, it does not rotate like a classical rigid object \cite{otsuka_2026}. 
The angle/angular-momentum uncertainty should be used cautiously, partly because the angle is periodic and  angular momentum is quantized more sparsely than linear momentum.
\\

\section{Fusion Reactions between Atomic Nuclei \label{fusion}} 

We here discuss relevance to the fusion reaction, where two nuclei touch and eventually fuse into one nucleus. We take an example of the reaction between the oxygen-16 ($^{16}$O, $Z$=8, $N$=8) nucleus and $^{154}$Sm (see Fig.~\ref{fig:fusion}), referring to the works\cite{hagino_2012,hagino_2022}.  
When these nuclei approach, their motion is decelerated by the Coulomb interaction, a repulsive force due to protons in both nuclei.  We can take a picture that the smaller nucleus the $^{16}$O enters into the larger nucleus $^{154}$Sm.  The $^{154}$Sm forms, around its surface, the so-called Coulomb barrier, which is 
 a bump of the effective potential between the two nuclei.  This barrier stops the 
 in-coming $^{16}$O nucleus in classical mechanics, if its kinetic energy is below the height of the barrier.  In quantum mechanics, however, the $^{16}$O nucleus can still penetrate into this barrier because of the tunneling effect.  The probability of the tunneling depends on the energy of $^{16}$O but also on the barrier height: the higher barrier, the lower probability.  Here, the shape of $^{154}$Sm matters.  

The shape of $^{154}$Sm is an ellipsoid discussed already (see Fig.~\ref{fig:overlap_ampl}), whereas the $^{16}$O is basically a smaller spherical nucleus.  They approach in various ways as shown in Fig.~\ref{fig:fusion}.  In Fig.~\ref{fig:fusion}{\bf a}, $^{16}$O approaches the edge of elongated part of $^{154}$Sm, where the barrier height is lowest compared to the other directions because of fewer protons nearby.  So, the tunneling starts to occur with a higher probability. The situation is opposite in Fig.~\ref{fig:fusion}{\bf c}, where $^{16}$O come to a side of $^{154}$Sm, and more protons nearby form a higher Coulomb barrier, reducing tunneling probability.
Figure~\ref{fig:fusion}{\bf b} indicates an intermediate situation.

Thus, the low-energy fusion reaction tends to occur through the process like Fig.~\ref{fig:fusion}{\bf a}.  The question is if the $^{154}$Sm nuclear state stays like Fig.~\ref{fig:fusion}{\bf a} until the tunneling proceeds sufficiently.  Here, the standing time of the intrinsic state of $^{154}$Sm becomes relevant.  Figure~\ref{fig:overlap_ampl} indicates that the intrinsic state in an initial direction remains so with 60 (20)\% probability at $t$=1 (2)$\times$10$^{-22}$ sec.
This number can be compared to the estimated time for completing tunneling, which has been shown to be about 6 $\times$ 10$^{-22}$ sec by Hagino\cite{hagino_priv} using WKB framework with a standard barrier curvature ($\sim$3.5 MeV)\cite{vaz_1978}. This comparison suggests that during the initial stage of the tunneling, $^{154}$Sm nucleus keeps its direction towards the incoming $^{16}$O, facilitating the fusion process. Once the tunneling enters the intermediate stage, the direction of $^{154}$Sm becomes less relevant.  We can thus understand the importance of deformation for fusion reaction and can verify, also from the present standpoint, the validity of the mechanism indicated by Hagino and Takigawa\cite{hagino_2012,hagino_2022}.  
\\


\begin{figure}[tb]
  \centering
\includegraphics[width=8.5cm]{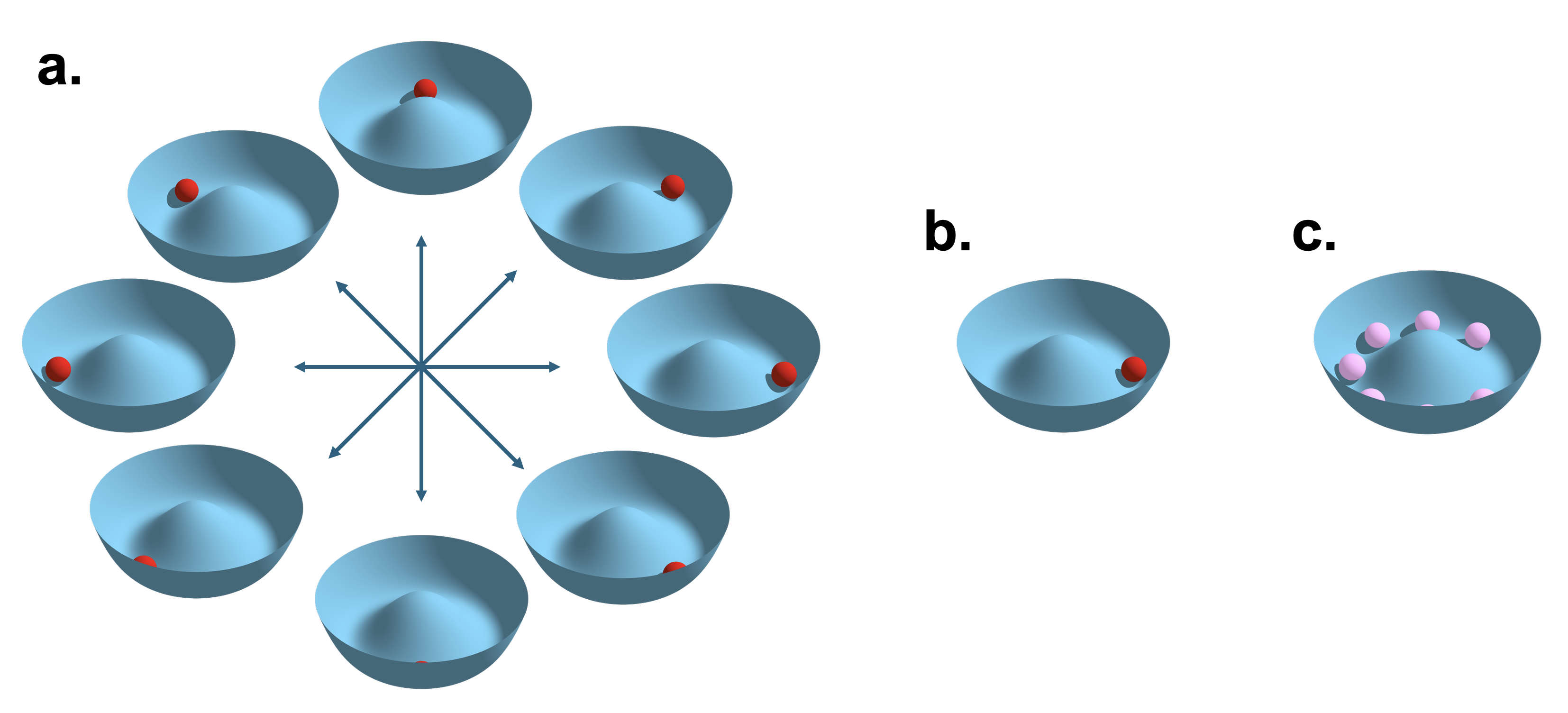}
    \caption{ {\bf Restoration of broken symmetry. }  
    {\bf a.} States in broken symmetry discretized into 8 directions.
  {\bf b.}  Intrinsic state at time $t$=0.
   {\bf c.} Dissolved states at time $t >$ 0 orthogonal to the one in panel  {\bf b.}.} 
  \label{fig:mex_hat}  
\end{figure}  

\section{Relation to Restoration of Broken Symmetry \label{broken}}

The rotational mode of atomic nuclei is created by the restoration of broken symmetry.
The symmetry in this work is the rotational symmetry, and eigenstates of atomic nuclei should have good angular momentum and its $z$-projection.  The intrinsic state $\phi_i$ does not fulfill this requirement, and breaks rotational symmetry.  

The broken symmetry is often described by a schematic Mexican-Hat potential and a ball in it, as shown in Fig.~\ref{fig:mex_hat}.  In the present case, a ball in the hat corresponds to the ellipsoid oriented in a direction.  
Figure~\ref{fig:mex_hat}{\bf a} indicates states of selected ball positions.  By some mechanism (nuclear Hamiltonian presently), the symmetry is brought in, and the eigenstate is given by the appropriate superposition of all ball locations.  A ball then takes one of those locations with a certain probability.  The present work can then be generalized: if the ball is like Fig.~\ref{fig:mex_hat}{\bf b} at time $t$=0, on what time scale this ball moves to any of pink ones in Fig.~\ref{fig:mex_hat}{\bf c}.  The probability that the red ball remains standing at its initial position is nothing but $|r(t)|^2$, and the system is in orthogonal states with the total probability $( 1- |r(t)|^2)$.

The present discussions imply that the time scale is basically determi{eq:phi}ned by the energy spectrum of the system, as modeled in eq.~(\ref{eq:St_time_est2}).  So the basic energy scale matters, and this feature is applicable for other general cases.


\section{Polymer (protein) \label{protein}}

As a possible application of the standing-time concept with a completely different time scale, we discuss the case of polymers.  Among various polymers, we take the example of protein. The protein is an assembly of molecules, and is known to be densely-packed with basically a rigid molecule configuration. So, it has good similarities on these aspects to atomic nuclei, from which one can expect the emergence of rotational modes.  

Some objects comprising many molecules are expected to exhibit rotational features.
If the relative configuration of the constituent molecules are sufficiently rigid, 
the $J(J+1)$ rule for rotational excitation energies with $J$ being the spin of the object is obtained within the formulation of the distant-object rotation mode (rotational mode of rigidly configured smaller objects like molecules) \cite{otsuka_2026}.  Note that the rotational modes of nuclei with ellipsoidal shapes, as discussed in previous sections, are treated as the mode of the compact-object rotation \cite{otsuka_2026}.  These two modes are formulated on the common mathematical procedure, but are driven by different parts of the Hamiltonian.
Thus, the argument for proteins can be made in parallel to that for ellipsodal nuclei presented in previous sections. The main difference is in the time scale or in the energy scale.    

In the distant-object rotation, even a spherical shape can produce rotational bands, as long as constituent molecules are rigidly configured in their relative positions \cite{otsuka_2026}. The moment of inertia, calculated from density distributions of the intrinsic state, are shown to be equal to the classical values at the limit of perfect rigidity \cite{otsuka_2026}.
While proteins exhibit a huge variety, we take a simple case with 500 amino residues with average molecular weight, 110, resulting in the total molecular weight, 55,000. We assume that these molecules are distributed around the center of gravity of the protein with the mean value of the squared radius, $\sim (10^{-9})^2$ \, m$^2$.  
The classical moment of inertia is estimated, in this case, to be $\sim$3 $\times$ 10$^{-11}$ MeV$\cdot$m$^2$/ $c^2$, where $c$ is the speed of light, and an isotropic uniform density profile is 
used as a simple approximation for the calculation of the moment of inertia.  
Among three rotation axes, one of them, called the $z$ axis, produces $K$=0 states with vanished rotational energies, and two other axes, $x$ and $y$, produce rotational energies, for which one is taken because of the $K$=0 superposition and the equal moment of inertia values \cite{otsuka_2026}.    

With this moment of inertia, the rotational excitation energy is given by about 2/3 10$^{-9}$ $J(J+1)$ eV = 2/3 $J(J+1)$ nano eV.  We thus expect rotational bands with an energy scale of 1-100 nano eV for $J$ up to 10. 

The relevant value of $J$ can be estimated by using the angle-angular-momentum uncertainty principle.
The size of amino acid is taken to be 0.5 nano m (n m = $10^{-9}$ m), and its distance from the center of the protein is taken to be 1 n m.  The angle of the amino acid viewed from the center of the protein, denoted by $\theta$, is estimated as ${\rm sin} \theta = 0.25 n m / 1 n m$, which results in $\theta \sim 0.25$. The uncertainty of the protein direction is considered not to exceed this angle.  By applying a simple uncertainty relation $\Delta \theta \times \Delta J \sim \hbar$, we obtain $\Delta J \sim 4$ with $\theta \sim 0.25$.  Because there are three axes, we obtain that $\Delta J \sim 7$, where $4 \times \sqrt{3} \sim 7$ is used.  This is just a simple estimate of the mean angular momentum of the intrinsic state of a protein with a near spherical configuration of Amino residues. Although there might be different estimates, we adopt it.  The value $J \sim 7$ produces an excitation energy $\sim$40 nano eV from the equation above.  This provides us with an estimate of $\Delta E_r (protein)$ within one band.   Considering contributions from some other bands, $\Delta E_r (protein) \sim 100$ nano eV is adopted hereafter.

Equation~(\ref{eq:St_time_est2}) is adjusted to the protein case, with $\Delta E_r$ in MeV replaced by $\Delta E_r (protein)$ in nano eV.  The resulting standing time is given by 
\begin{equation}
\tau_s \, \sim \, 6.6 \, \eta \, \frac{100}{\Delta E_r (protein)} \, {\rm nano \,sec}, 
\label{eq:protein}
\end{equation}
where 1 nano sec = $10^{-9}$ sec.
By adopting $\eta \sim 0.25$ based on the argument for eq.~(\ref{eq:St_time_est2}) and using $\Delta E_r (protein) \sim 100$ nano eV mentioned above, we end up with the standing time for the protein, $\tau_s \, \sim$2 nano sec with possible variations of a similar scale.
This result can be written as $\tau_s \Delta E_r \sim \eta \hbar$, similarly to Sec.~\ref{standing}. 
 
If this standing time can be observed by cooling a single protein polymer to almost zero temperature in a vacuum  environment with no external force (like in space), it will be of great interest for the study of protein and also for quantum mechanics. It needs an innovative idea to detect the direction of the protein without disturbing it. Another interesting dream is to accelerate a protein molecule (possibly by laser) and collide it on a proper target (including the same protein molecule), as is done for atomic nuclei with the RHC (see Sec.~\ref{concrete}).

In such ways faster than the present standing time, we might  be able to extract some structural information of the protein molecule.

\section{Electron Drop in a Hydrogen-type potential \label{electron}}

It is of interest to apply this framework to an electron placed and bound in a Hydrogen-like atom.  
This case has nothing to do with shapes, but involves the time evolution and the energy scale. 

An electron is placed into the atomic Coulomb potential with negative energy.  Suppose that it is at a location $\vec{x}$ at time $t$=0, where the electron may be a tiny wave packet, i.e., a drop. This initial wave function is expanded by Coulomb wave functions which have energy eigenvalues <0. As the energies of those wave functions are all negative with a lower boundary also, the energy is confined. The standing time can then be finite. By setting the appropriate energy scale $\sim$eV, the time scale may become $\sim$10$^{-15}$ sec=1 femto sec. It is of interest how one could put an electron into a location with a negative energy in an atom.  This is different from a free particle, which can have only vanished standing time (i.e., very short for a wave packet), because a free electron at a location is expanded by wave functions going into infinitely high energy.   There could be analogous situations, which are extremely intriguing. 
\\

\section{Fission and $\alpha$ Decay/Emission \label{fission}}

\begin{figure}[tb]
  \centering
\includegraphics[width=7.0cm]{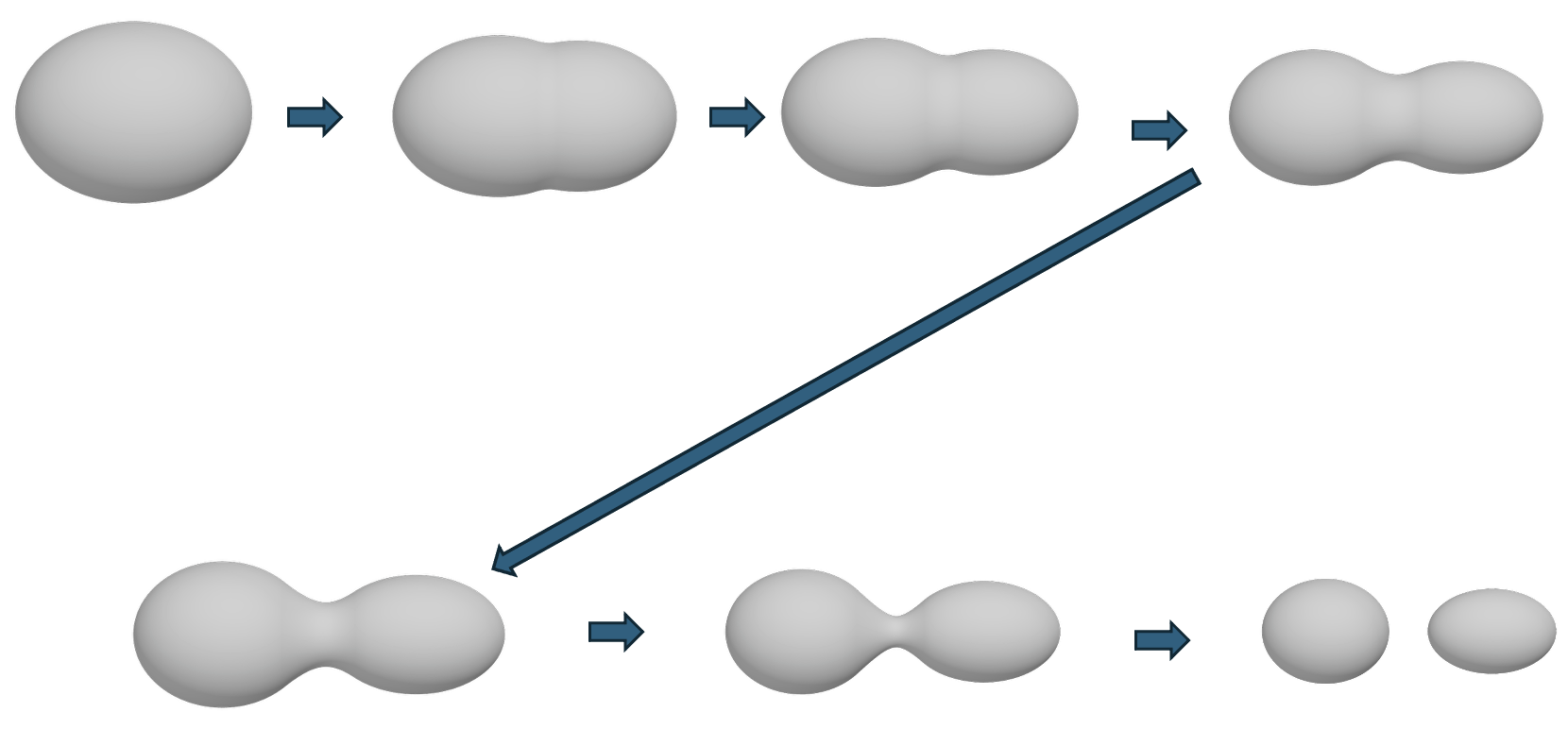}
    \caption{ {\bf Schematic illustration of fission process. }  
    Arrows indicate the fission process as time proceeds.
 }
  \label{fig:fission}  
\end{figure}

The fission process also shows interesting features related to the standing time.
Once the shape of the fissioning nucleus approaches the scission as shown in Fig.~\ref{fig:fission}, rotational excitation energies in a metastable picture become lower, likely increasing the standing time.  If the fission process reaches around the scission point, the rotational scheme, still in a metastable picture, will change from the compact-object (e.g., ellipsoid) rotation to the distant-object (e.g., molecule) rotation\cite{otsuka_2026}, and the rotational excitation energies may become further lower.
As shown in eq.~(\ref{eq:St_time_est2}), this implies that the standing time becomes longer and longer.  If the standing time is prolonged to be comparable to typical time scale of the phenomena of interest, e.g., fission, the situation becomes, in its appearance, similar to the spontaneous symmetry breaking. 
In this view, the fission process may be discussed without referring to rotational symmetry, in a reasonable approximation.  
The time scale of fission is suggested\cite{hinde_1992,ichikawa_2012,scamps_2018, schunk_2022} as 10$^{-19}$-10$^{-21}$ sec.
The standing time depends on the representative energy shift $\Delta E_r$, which is likely smaller by an order of magnitude for fissioning nuclei than the values discussed so far for rare-earth nuclei, because of heavier mass and more elongated ellipsoids.   The standing time is then expected to be longer than 10$^{-21}$ sec for each stage of the whole fission process.  This estimate can be compared to the time scales shown above, with possibly new insights.

The standing-time argument may also be applied to $\alpha$ decay\cite{alpha_decay_wiki} and reaction-induced $\alpha$ emission, which have some common aspects with fission. Because the formation of $\alpha$(-like four-nucleon system) + the rest of the nucleus, which is a virtual excitation from the ground-state wave function in conventional structure theories such as the shell model, implies a coupled deformed object, a longer standing time (in the present time scale) is generally expected for this meta-stable formation: The barrier height becomes lower for an outgoing $\alpha$ due to the elongation, and its longest axis remains during this standing time, which can help outward motion of $\alpha$, beyond the simple pre-formation estimate\cite{tanaka_2021}. These properties seem to enhance certain cases of $\alpha$ decay and reaction-induced $\alpha$ emission. Further studies along this line may be of great interest.
\\


\section{Fusion for Superheavy Element Synthesis \label{super}}

The prolongation of the standing time with strongly deformed nuclei may be utilized or are already used implicitly for the synthesis of superheavy elements\cite{karpov_2013,erler_2012}, in the essentially same manner as those for fusion and fission discussed so far.  
In general, the fusion reaction is a possible way for the synthesis of superheavy elements\cite{smits_2024,oganessian_2024}.
The present case corresponds to hot fusion\cite{tanaka_2018,tanaka_2020}, compared to cold fusion where fusing nuclei are not strongly deformed.
By choosing strongly deformed nuclei with low energy scale, one may increase the standing time, and then enhance fusion reaction or something similar, as discussed for the fusion between $^{16}$O and $^{154}$Sm nuclei.  More concrete details will be a productive future project. 
\\


\section{Summary and Remarks}

We showed some features arising from the time evolution of an ellipsoid-shape intrinsic state as a component of an eigenstate with good $J$.  This time-evolved state remains almost unchanged for a certain time period called the standing time.  This consequence naturally arises in the line of the fully quantum-mechanical formulation of rotational modes in nuclei\cite{otsuka_2025,otsuka_2025a,otsuka_2026}, where the rotational symmetry is fully  incorporated but the classical image of the rigid-body rotation is absent.  In other words, this is an explicitly time-dependent property given by stationary eigenstates. Fusion enhancement at low energy is consistent with this feature.

One can therefore perform an experiment on this intrinsic state, provided that the relevant phenomena take place within the standing-time scale.
This provides us with support to the use of RHC for extracting certain properties of intrinsic states of deformed nuclei, as the standing time is much longer than the interaction time of two nuclei in the RHC.
This argument justifies the use of an intrinsic-state picture for both conventional one-body density approach\cite{giacalone_2022,bally_2022,jia_2022,dimri_2023,zhang_2022,Ryssens_Giacalone_2023} and more recent two-body density approach\cite{duguet_2025,blaizot_2025,bofos_2026}.  If the intrinsic state is substantially replaced by those in different directions due to time evolution during the collision, the extraction of shape information would be difficult, if not impossible.  
In this respect, the present results clarify basic validity of the RHC approach.

We simply followed the time-dependent solution of the Schr\"odinger equation. It is then natural that the crucial factor for the standing time is energy scale.  
For instance, in the case of nuclear shapes, shape parameters or their fluctuations are not directly related to the standing time but can be indirectly so through energy spectra. 
The major role of the energy spectrum may be useful, when the present features are explored in a variety of cases and systems.
Indeed, the concept of the standing time shows varying relevances and values in fusion and fission reactions, including hot fusion for superheavy element synthesis.  Intriguing features emerge for a protein molecule and an electron bound in an atom.

The standing time for heavy rare-earth nuclei is much shorter than the time scales of electro-magnetic transitions even down to E1 or M1 transitions\cite{bohr_mottelson_book1}.  Thus, electromagnetic and RHC experiments are complementary. The former type of experiments produce data for physical quantities after the superpositions of intrinsic states (see Fig.~\ref{fig:rot_state}{\bf b}). While those data reflect the intrinsic shapes, what matters is their sensitivity to certain properties, such as triaxiality or clustering. 
The latter method provides what the former one cannot, but is applicable to ground-state properties only.
Thus, both approaches are useful to firmly establish finer and modern picture of nuclear shapes.

The present time scale differs, by a factor of 100-1000, from that reported in an attempt using the uncertainty principle between angle and angular momentum\cite{nakatsukasa_2016}. 
This estimate was mentioned in \cite{STAR_2024}, which 
sparked a commentary warning paper stating no relevant time scale for the issue \cite{doba_2025}. 
Our analysis goes beyond this commentary arguments of \cite{doba_2025}, unveiling that there is an additional quantum mechanical feature underneath, resulting in the present time scale, which is a specific case of the general and broader time-evolution picture. 

Another related subject is the nuclear-size measurement by (relatively) high-energy heavy-ion collisions with the eikonal approximation\cite{tanihata_2013,tanaka_2024}. If touching between projectile and target nuclei is completed, more or less, within a time period comparable to the standing time of the deformed one of two nuclei, the process may vary for different orientations of this deformed nucleus upon the contact, and certain shape-sensitive properties may be extracted besides those representing the radius as usual. 
\\

\noindent {\bf Acknowledgements}
The authors appreciate very much Prof. R. F. Casten for valuable advices and careful reading of the manuscript. 
T.O. thanks Prof. Y. Aritomo for valuable discussions on fission reactions, Prof. K. Hagino for valuable results and discussions on fusion reactions, and Prof. A. Poves for useful discussions.
T.O. is sincerely grateful to Dr. T. Kobori for valuable contributions and stimulating relevant comments. 
This work was supported in part by MEXT KAKENHI Grant No. JP25K00998.
\\ 
  

\appendix
\section{Rotational Modes in Quantum Mechanical Formulation}

We present a quantum mechanical formulation of rotational modes, that is free from the classical (or semi-classical) picture/interpretation where a deformed nucleus (or object) is regarded as an axially-symmetric rigid-body literally rotating.
A rotational band in the present formulation can be defined as a set of many-nucleon (or particle) states, where the member of angular momentum $J$ is generated by projecting a common intrinsic state onto this angular momentum $J$.  This definition itself may not be new, but we briefly re-formulate below the whole description of rotational bands, staying inside quantum mechanics (without the image of dynamically rotating axially-symmetric rigid-body, as widely supposed in the past). 

\begin{figure}[hbt]
\begin{center} 
\includegraphics[width=6cm]{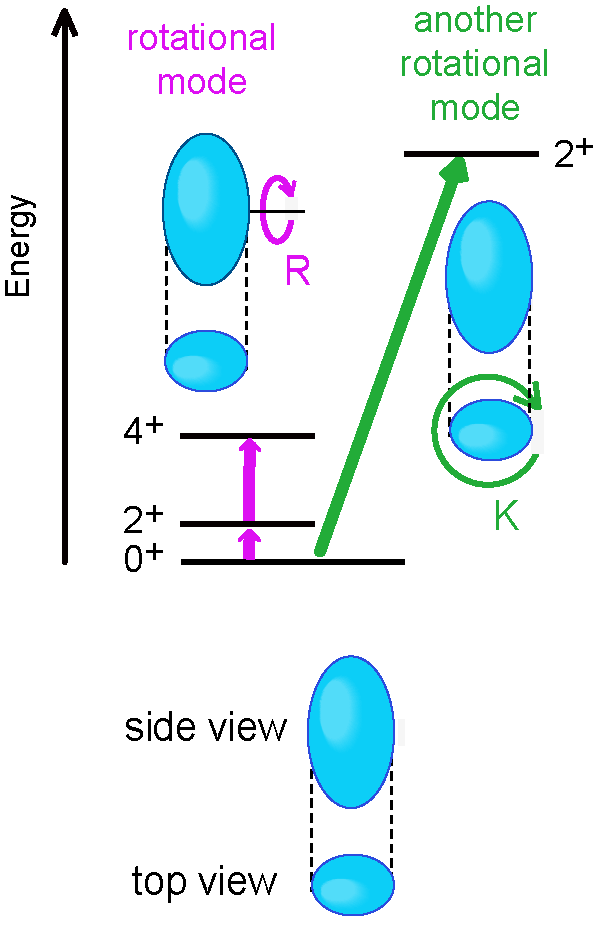}
  \caption{{\bf  Two rotational modes.}  
  Two rotational modes of nuclei in an ellipsoidal shape with triaxiality are shown.  One occurs about the axis perpendicular to the longest axis of the ellipsoid, and the rotational mode is denoted by R in purple color.
  The other rotational mode is about the longest ellipsoid axis which is in parallel to the direction of the top view, and is denoted by K in green color.
  } 
\label{fig:rotational mode}
\end{center}
\end{figure}

We use the angular-momentum projection method\cite{ring_schuck_book} formulated with Wigner's $D$ function\cite{edmonds}, beginning with its concise sketch.  First, the intrinsic state is denoted by $\phi_i$ as in the main text.  The state $\phi_i$ represents a quantum-many body state with an ellipsoidal shape schematically shown in the lower part of Fig.~\ref{fig:rotational mode}.  It can be a sophisticated state containing full of correlations by nuclear forces.  So, it does not have to be a simple state.  From this $\phi_i$, we obtain the state of definite $J$ and $M$, the total angular momentum and its z-projection in the laboratory frame.  This projection can be performed by rotating $\phi_i$ in the three-dimensional space with three Euler angles $\alpha, \beta$ and $\gamma$, and by integrating it with an appropriate weighting factor, Wigner's D function\cite{edmonds}.  The obtained state is written as,
\begin{eqnarray}
&\Psi \bigl[\phi, J, M, K \bigr] \, = \, \frac{2J+1}{8 \pi^2} \, \int_0^{2\pi}   d\alpha \int_0^{\pi} d\beta \, {\rm sin}\beta \, \int_0^{2\pi} d\gamma \,\,\,\,\,\,\,\,\,\,\,\,\,\,\,\,\,\,\, \nonumber \\
& \,\,\,\,\,\,\,\,\,\,\,\,\,\,\,\,\,\,\,\,\,\,\big\{ D^J_{M,K} (\alpha, \beta, \gamma) \big\}^*\, e^{i\alpha \hat{J}_z} \, e^{i\beta \hat{J}_y} \, e^{i\gamma \hat{J}_z} \, | \, \phi_i \rangle, 
\label{eq:rot_D}
\end{eqnarray}
where $D$ is the Wigner's function.   For more details, for instance, see eq.(9) of \cite{otsuka_2025} where \cite{ring_schuck_book} is cited.

Equation~(\ref{eq:rot_D}) implies that the three-fold rotation of $\phi_i$ generates states with good ($J$, $M$) pairs.
One notices an additional index of $K$.  In fact, there can be different and independent states from the same $\phi_i$ for a given pair ($J$, $M$), and $K$ specifies them. 
By performing the $J_z$ rotation ($e^{i\gamma \hat{J}_z}$) with proper weighting factor, 
we project $\phi_i$ onto a specific value of $K$, the $z$-component of $\vec{J}$ of $\phi_i$.  
The intuitive image of $K$ is given in Fig.~\ref{fig:rotational mode}.  

Multiple band structure is classified by $K$, as individual bands have almost pure values of $K$ as demonstrated in \cite{otsuka_2025}.  Namely, $K$ is a practically good quantum number for strongly deformed nuclei with triaxiality\cite{otsuka_2025}.  The triaxiality appears in virtually all heavy deformed nuclei\cite{otsuka_2025,otsuka_2025a}. The lowest band in energy is the $K$=0 band, which is nothing but the band on the left-hand side of Fig.~\ref{fig:rot_state}{\bf c}.

In the time-evolution calculations described in the main text, $K$ value is taken from $K$=0 up to $K$=20.  Because of the symmetry of ellipsoidal shape, only even values of $K$ appear. 
Contributions to the present time evolution from $K >$ 8 are really small.  
Other practical setup is indicated in the main text.\\



\vspace{1cm}

\noindent {\bf Data Availability} 

\noindent
All data relevant to this study are shown in the paper, but if more details are needed, they are available from the corresponding author upon reasonable request.
\\

\noindent {\bf Code Availability} 

\noindent
There are no special codes involved in this work.
\\

\noindent {\bf Author Contributions}
\noindent
All authors made contributions indistinguishable.
\\

\noindent
\textbf{Competing interests} \\
\noindent
The authors declare no competing interests.
\\\\
\noindent
\textbf{Correspondence and requests for materials} should be addressed to T.O.


\makeatletter
\renewcommand\@biblabel[1]{#1.}
\makeatother

\end{document}